\documentclass[12pt,aps,amsmath,amssymb,sort&compress,comma,super,nofootinbib]{revtex4-1}
\usepackage{graphicx}
\usepackage{epstopdf}
\usepackage{subfigure}  
\usepackage{setspace}
\begin{document}

\newcommand{\mnras}{MNRAS\/}
\newcommand{\araa}{ARA\&A\/}
\newcommand{\aap}{A\&A\/}
\newcommand{\apjl}{ApJ\/}
\newcommand{\aj}{AJ\/}
\newcommand{\etal}{{\it et al.}\/}
\newcommand{\gtwid}{\mathrel{\raise.3ex\hbox{$>$\kern-.75em\lower1ex\hbox{$\sim$}}}}
\newcommand{\ltwid}{\mathrel{\raise.3ex\hbox{$<$\kern-.75em\lower1ex\hbox{$\sim$}}}}

\title{Active Galactic Nuclei and Quasars: Why Still a Puzzle after 50 years?}

\author{Robert Antonucci}
\affiliation{University of California, Physics Department, Santa Barbara, CA 93106-9530, USA}


\begin{abstract}

\begin{quote}
``The history of scientific and technical discovery teaches us that the human
race is poor in independent thinking and creative imagination." -A Einstein
\end{quote}

The first part of this article is a historical and physical introduction to
quasars and their close cousins, called Active Galactic Nuclei (AGN).  In the
second part, I argue that our progress in understanding them has been
unsatisfactory and in fact somewhat illusory since their discovery fifty years
ago, and that much of the reason is a pervasive lack of critical thinking in
the research community.  It would be very surprising if other fields do not
suffer similar failings.

\end{abstract}


\maketitle
\tableofcontents

\vfill\eject
\section{Early observations and physical inferences}

Quasars were discovered by M. Schmidt in 1963, so this is approximately their
fiftieth anniversary.  They are extremely powerful unresolved sources
of optical/UV
light.  Specifically, quasars produce up to $\sim1\times10^{39-40}$ J/s,
which is $\sim1\times10^{13}$ times
the luminosity of the sun, and $\sim1000$ times the luminosity of our entire Milky
Way galaxy, which contains $\sim1\times10^{11-12}$ stars.  Quasars also emit X-rays, which are
highly variable and show fascinating atomic emission features.

About 10\% of them (called ``radio-loud") are in addition powerful emitters of
radio waves.  The radio waves are produced by the synchrotron process, that is,
they come from relativistic electrons spiraling in magnetic fields.  The
radio emission is spatially resolved (e.g., Fig.~2), and usually takes the form of two huge
($\sim100$kpc or 300,000 light years each) gorgeous ``lobes" situated on
either side of the optical nucleus.
\begin{figure}[htbp]
\includegraphics[width=7.5cm]{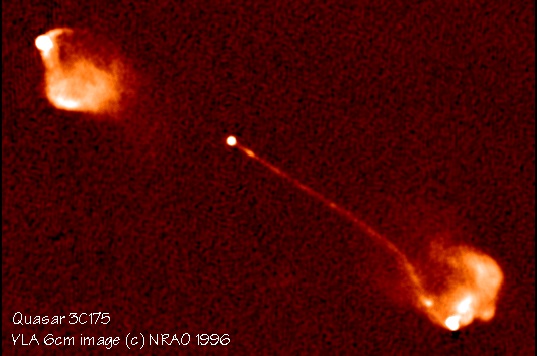}
\caption{A gorgeous VLA map of a radio loud quasar, showing the parsec-scale core,
the (apparently) one-sided jet, and the double lobes. Credit: National Radio
Astronomy Observatory.
\label{fig:1}}
\end{figure}
Linear features called jets connect the tiny (parsec-scale) radio cores to
the lobes, feeding
them energy in the form of relativistic electrons and magnetic field.  But the
jets often appear one sided!  How is the other lobe energized\dots?

Certain radio loud quasars showed only point sources on early maps
with $\sim1$ arcsec
resolution.  These tiny (pc-scale, milliarcsec on the sky) radio cores
could however be mapped with Very Long Baseline
Interferometry, and they almost always turned out to show an even
tinier (sub-pc) stationary point
source, with a line of little blobs flying outwards along one side like
cannonballs.  This tiny linear feature is just the base of the larger-scale
($\sim10$--100kpc) jet.  The blobs appear to move perpendicular to the line of sight,
often at about ten times light speed (this is called superluminal
motion)!  That would
be very, very verboten in relativity.  We also now know that quasar
nuclei are at
the centers of galaxies, primarily nascent galaxies in the early universe.

I wish to recount the history of quasars and the closely related objects called
Active Galactic Nuclei (AGN), in such a way as to bring in the key observations
and physical ideas.  The term AGN has historically been used for 1) Seyfert
galaxies (1940s) --- which turn out to be just weaker versions of the ``radio
quiet" quasars.  2) radio galaxies, discovered in the 1950s.  Many are just
the lower-luminosity versions of radio-loud quasars.  Today the
quasars themselves
are often grouped under the AGN rubric.

When the giant radio galaxies were discovered, two inferences were quickly
made.  One is that the Steady State theory of the universe (eternal,
unchanging) was no longer tenable, because we observe a large excess of sources
at faint flux levels relative to extrapolations from the bright sources.
Roughly speaking, this requires that as we look far back in time towards
the Big Bang, their space density increases, and thus the universe evolves!

The second inference is that the energy content of the radio lobes is
astonishingly high.  Suppose we want to make a very luminous synchrotron
source, using as little energy as possible.  (Suppose god is cheap.)
To do this, we would put
(approximately) equal amounts of energy into relativistic electrons and into
magnetic field.  So this case provides a lower limit to the amount of energy
contained in the lobes, given their observed luminosity.  The results are up to
$1\times10^{54-55}$ J, which by the equation $E=mc^2$ is the mass equivalent of about ten million
stars like the sun!  That is, one would have to (hypothetically)
annihilate millions
of stars and anti-stars to produce such energies.  Where did this
energy come from?

Some smart theorists, most notably D.~Lynden-Bell in 1969, proposed
that the energy
production must involve gravitational collapse of hugely massive gas clouds
($1\times10^6$--$1\times10^9$ solar masses) to relativistic (near black-hole sized) dimensions.
The energy available from gravitational collapse is extremely sensitive
to the compactness of the final mass distribution.  Collapse to relativistic
dimensions results in almost a zero-divide in the potential energy released ---
enough to power the quasars.  (This effect accounts for most of the ``fireworks"
throughout astronomy.)  Furthermore the required black holes themselves would be
only solar-system sized, consistent with the extremely small size upper limits
based on rapid optical/UV variability, together with causality.  That is, they
are about one billionth the size of a galaxy.

I will only discuss the relativistic regions of AGN in this essay.

\section{What are the main observations and theories required to confirm and
physically understand this scenario?}

Here is my personal and very incomplete list.

\begin{itemize}
\item Do we see the required remnant (starved) black holes in the centers of
nearby (present-day) galaxies, left over from the prime quasar era when the
universe was $\sim20$\% of its present age?

\item How, specifically, is the prodigious electromagnetic luminosity produced by
gravitational collapse or infall?

\item We find that very powerful radio (and optical, and X-ray)
synchrotron-emitting plasma jets emerge from the central sub-parsec cores of
some of them --- but mysteriously, just the ones which reside in elliptical
galaxy hosts!  Why is that?  Why should an engine of solar-system size care
what type of galaxy it resides in?

\item Why do these cores shoot ``cannonballs" of plasma, often at apparently
superluminal speeds, and apparently on only {\it one side} of the cores?  The plasma
jets feed the giant and extraordinarily energy-rich radio lobes.  How does the
bulk kinetic energy of the jet plasma get thermalized to produce the relaxed-looking giant lobes?

\item How can we probe spacetime close to the putative black hole --- can we prove
that the Schwarzschild or Kerr solutions of Einstein's equation for the
spacetime geometry around a black hole are correct?

\item We are only now becoming aware of the ecological role of the quasars'
immense radiative and mechanical luminosity in the formation of galaxies and
stars.  The probable importance of this AGN ``feedback" became undeniable
in the last 20 years when close
connections were shown between the central black hole masses and the
stellar content
and internal (stellar) orbital velocities.  And as noted,
only those black holes that live in Elliptical (not Spiral) galaxies have the
blockbuster radio power.  Host galaxies won't be discussed further in this
article --- but as a teaser, the quasar momentum and energy input may block
further mass accretion onto a protogalaxy, and may in fact blow the
interstellar gas out of the galaxy body, quenching star formation, and setting
the maximum mass of (ordinary, ``baryonic") matter for these universal building
blocks.
\end{itemize}

Only some of these great mysteries can be elaborated below.

\section{The ``breakthroughs," robust, dubious, and falsified}

I will now enumerate the observations and interpretations that seem to most of
the community to be the pillars of our understanding of AGN physics.  In
my opinion, several of these inferences are robust, others are dubious,
and still others have been
fully falsified, yet continue to be used by many researchers who
seem to lack a good critical faculty, and who thus waste vast amounts of time
and resources.

Many theory papers have already been ruled out by observations
by the time they are published.  Observers routinely use models to interpret
their data long after the models have been falsified.

\subsection{Prediction of Leftover Supermassive Black Holes: {\it Robust
Confirmation}, 1980s-1990s.}

An essential prediction of any gravitational collapse model is that leftover
(starved) supermassive black holes (or other tiny objects) reside at
the centers of most normal
galaxies in the present universe, and this has been spectacularly verified,
e.g. Kormendy 1988.

\subsection{The Unified Model, Part 1:  Relativistic beaming: {\it Robust
Confirmation}, 1980s}

Up until about 1980 or so, AGN were divided into many puzzling
phenomenological subtypes
based on correlated suites of observed traits.   Much of the confusion was
cleared up in the following decade:  it turns out that while we see
dramatically varied
behavior, most of the differences depend only on the inclination of these roughly
axisymmetric sources to the line of sight to Earth!
\begin{figure}[htbp]
\includegraphics[width=7.5cm]{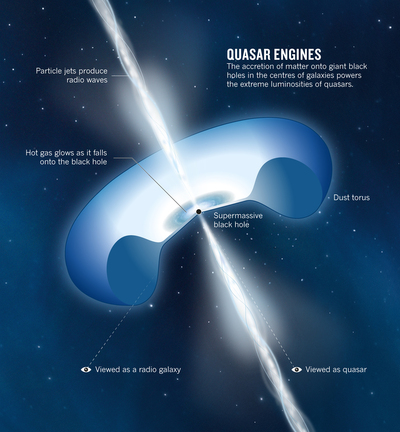}
\caption{Illustration of the central parsecs of AGN. Shown are the Supermassive
Black Hole, the accretion flow believed to be feeding the black hole and producing
thermal optical/UV emission, the obscuring torus, and the jets, which are only
strong in radio loud objects. The ``torus" extends to the sublimation radius and
its obscuration is primarily equitorial, but no implication is made regarding its
actual shape and outer extent. (From Antonucci {\it Nature} {\bf 495}, 165A, 2013)
\label{fig:2}}
\end{figure}
This isn't so surprising in retrospect:  people have many systematic
differences in
appearance, depending on whether they are seen from the front or from the top.
(The effect is relatively unimportant for stars because they are round.)

The superluminal speeds are now robustly attributed to motions actually at
{\it nearly} light speed, traveling roughly (but not exactly) towards us along our
line of sight.  This is a classical effect due to sequentially lower light
travel times for the little cannonball components shooting out of the radio
cores as they move closer to us, producing the {\it appearance} of faster than light
motions perpendicular to the line of sight.  These sources are picked up
preferentially because of the beaming (``headlight") effect that comes from
Special Relativity.  In fact, this beaming effect nicely explains the
apparently one-sided jets:  we are quite sure now that there {\it do} exist
``counterjets" so that both lobes are being energized, but the jet on the far
side beams its radiation away from the line of sight! For a detailed
observational review, see Antonucci 1993.

A spectacular corollary of the beaming idea is that there must be a much larger number of
equivalent radio sources whose jets are {\it not} pointed at Earth.  We now know that in
most cases the misdirected objects are none other than the normal giant double
sources (radio galaxies and the majority of radio quasars), which do {\it not} show fast
superluminal motion as seen from Earth. The most conclusive evidence for the latter
statement is that deep radio images of superluminal sources show large scale diffuse
and isotropically emitting radio components as well as the tiny beamed cores. We
must be able to see this diffuse emission in the misdirected objects, and only
normal doubles have the right properties. Credit goes mostly to Peter Scheuer, Roger Blandford,
Martin Rees, and Mitchell Begelman on the theoretical side.  If you are a specialist
and you think that a lot of progress has been made in AGN radio astronomy in the last
thirty years though, you'd find it interesting to consult the long review by Begelman
Blandford and Rees from 1984.  The open questions of jet launching, acceleration, confinement,
composition, equipartition, proton energies, and filling factors are still
basically with us.

\subsection{The Unified Model, Part 2:  Hidden Nuclei: {\it Discovery and
Robust Confirmation}, 1980s and 1990s.}

Most optical spectra of both radio loud and radio quiet AGN come in two types
(called Type 1 and Type 2!).  The Type 2 sources have apparently simpler spectra,
showing just a weak nuclear continuum which is constant in time, and often spatially resolved, plus a
set of emission line clouds (the ``Narrow Line Region") with velocity
widths of a few hundred km/s, and with densities more in the range of
$1\times 10^3$--$1\times10^5$ H atoms per cubic cm.
These lines differ from those of H II regions in that they show much stronger
high ionization lines, and also stronger very low
ionization lines, compared with those of intermediate ionization. The difference
is due to the extremely broadband exciting continuua in AGN relative to 
the stars that power H II regions.

Type 1 AGN show the exact same NLR, but much more.  One also sees a
strong unresolved and variable central
source, including the energetically dominant ``thermal"
radiation component, or Big Blue Bump, in the optical/UV region, and
also a collection of ionized
gas clouds (the ``Broad Line Region") with collective Doppler-broadened
line profile widths
of a few thousand km/s.  Certain line ratios prove that these clouds
are relatively
dense, $\sim1 \times 10^{10}$ H atoms per cubic cm.  The Type 2 sources were
thought to lack these essential nuclear components, and thus to
differ fundamentally from the Type 1s.

My former thesis adviser Joseph Miller and I were able to sort this out by
separating out a trace of polarized (reflected --- as off your car windshield)
light in the Type 2s from the much stronger direct light from stellar and gas
emission.  We used a natural gaseous and sometimes dusty ``periscope" to see
the hidden central regions of
the Type 2 AGN ``from above."  To our astonishment and delight, they look
exactly like the ordinary directly visible Type 1 nuclei.  The Type 2s must then
just be unfavorably oriented from Earth's point of view in that obscuring
clouds of dusty gas lie in the line of sight.  The scattering polarization
position angle requires the obscuring material preventing our direct view
(as opposed to the scattering polar ``mirror" gas) to form (crudely) opaque
tori of dusty gas whose axes are parallel
to the radio structures, when the latter are detected.
There is a very important caveat to all this: while the most luminous radio
galaxies are all almost certainly quasars hidden by dusty tori, at lower levels
of radio luminosity, many objects intrinsically lack the characteristic powerful
optical/UV ``thermal" continuum. Many arguments from observations at all
wavelengths are reviewed in detail in Antonucci 2012. Also claims for the existence
of ``True Seyferts 2s," said to behave like Type 1 objects except that they lack
broad line regions intrinsically, are critiqued in detail in Section 2 of that paper.
Most are found wanting.

\subsection{The putative quasi-static accretion disk of inflowing matter.
{\it Falsified}, 1980s--present.}

Here we are discussing the ``thermal" AGN such as quasars (Antonucci 2012),
in particular the optical/UV Big Blue Bump
component of their spectral energy distribution.  This component is
energetically dominant and nearly
universally interpreted as optically thick thermal
radiation which arises in the region of steeply falling potential region near
the black hole.

``Wouldn't It Be Loverly" (from My Fair Lady) if\dots as accreting matter spirals
down the gravitational potential well, it radiates the incrementally released
gravitational potential energy right where it's produced?  (I'm
ignoring a couple
of unimportant subtleties\dots)  The region where (effectively) most of
the potential drop occurs
has a size of a few to $\sim10$s of times the event horizon radius, which for a
non-rotating black hole is given by the formula 3 km times the black hole mass
in terms of the Sun's mass.  This idea, together with the key assumption of a
``quasistatic" flow in which the inflow timescale is much larger than the other
timescales in the problem --- roughly but robustly --- predicts a certain spectral
energy distribution (amount of light at each frequency across the
electromagnetic spectrum).  Alas, this prediction was falsified almost
immediately
after the model was first used (Shields 1978, Malkan 1985).   A key
here was the realization that the infrared
radiation is from hot dust, and it cannot be extrapolated under the
optical spectrum
to help with the fitting as was done in the early disk papers.

On the other end of the spectrum, the
standard model predicts an exponential
falloff from the Wien part of the quasi blackbody radiation from the
innermost disk annulus,
which is never seen, but can be hidden with plausible but ad hoc Comptonization.

There are two possible candidates for some type of spectral feature from
the inner disk edge.  Many sources show a puzzling break (slope
steepening) below $\sim1000$\AA.  Though not an exponential, this might conceivably be
identified with an inner disk temperature.  Unfortunately,  according
to the models, it can't be fixed
in wavelength as observed; it must depend on black hole mass,
Eddington ratio, and
spin.
Similarly, there is a generic ``Soft X-ray Excess" in thermal AGN such as
quasars.  It's not clear whether this should be considered an
extension of the Big Blue Bump, but it's energy ($\sim200$eV) is again the same for
all objects, a fatal flaw for the standard disk model and almost all
of its variants.

Falsifications of other robust predictions soon followed.   In fact
they were already implicitly falsified by existing data.
For example, AGN are highly variable.
The region in which the bulk of the radiation from a standard accretion disk
arises is fixed over human timescales because it's set by the black hole mass.
Therefore it must show a temperature increase roughly in proportion to 
$L^{1/4}$ which is irreconcilable with the observations (e.g. Ruan et al 2014).
This prediction is based on two extremely mild assumptions:  the validity of the
Stefan-Boltzmann Law (the luminosity per square meter of a blackbody radiator is
proportional to temperature to the fourth power), and the size of the event
horizon radius (and perhaps the innermost stable circular orbit) from
general relativity.  As a quasar continuum luminosity is
seen to vary (necessarily in this scenario, it's radiated from a fixed
area), it follows that one should be able to fit a temperature for the inner
edge of such an accretion disk, and it should vary according to T proportional
to $L^{1/4}$.    This behavior has been seen in the accretion disks
inferred for black hole binaries! In most [but not all!] AGN, the continuum does become
bluer, but the frequently seen slope change remains at the generic rest wavelength
of 1000\AA, as does the Soft X-ray Excess.

There have been a few critiques of these models over the years, including
Antonucci et al 1989, 1999; Courvoisier and Clavel 1991; and Blaes 2007.
Most theorists now acknowledge
that the proposed workarounds are themselves seriously problematic.

These and other demurs were roundly and uncritically dismissed for
decades (and currently!) by almost the entire community, and enormous
effort has been spent refining this erroneous model.  This is a powerful example
of the wheel-spinning in our community which prevents us from making
rapid progress.

The theoretically essential quasi-static assumption was falsified explicitly
when it was shown that the optical continuum varies closely in phase with the
hydrogen-ionizing continuum (photons with energies above 1 Rydberg);
see Alloin et al 1985; Krolik et al 1991.  But we already knew enough about quasar
variability at that time to tell us we were
on the wrong track.

Might the optical/UV emitted still be some kind of chaotic blackbody disk of
indeterminate physics, but which somehow mimics the quasistatic disk
morphologically? Even this hope has been falsified. Recently ($\sim2000$'s) the
angular size-measuring technique called gravitational microlensing has produced
approximate but consistent and reliable source sizes for these heretofore
emitting regions, and they are typically several times as large as they can
possibly be in any standard disk models!\footnote{The large sizes from
microlensing are confirmed in some cases by reverberation time delays: McHardy
et al.\/ 2014 and references therein. Recently a particularly robust confirmation
was reported by Edelson et al. 2015, despite the reverberation bias toward short
lags and the assumption of zero rotation.}
(The sizes are quite approximate in individual cases, but the whole data set
together is compelling.)  That is, the surface brightness of the
optical/UV radiator
is {\it only a few percent} of that of a blackbody or any disk that fits the
spectra slopes locally. {\it This is incredibly important because it means
the radiation doesn't come from the region in which the energy is released.}
I conclude from this that we know next to nothing about the fundamental
physics of radiation from AGN.

I can't take a model seriously if it doesn't produce a surface brightness
of the right order of magnitude. Dexter and Agol (2011)
presented a toy model for a disk which is mostly dark at any particular
wavelength, but has bright (atypically hot) spots;  that way the
overall size of the
source could be increased
to match the observations.  This seems promising because polarization
data do indicate that optically thick emission
makes the Big Blue Bump  (Kishimoto et al 2004, 2008). Few if any
other theorists have even tried to build
the game-changing microlensing sizes into their models.

Recently two theory groups proposed that the standard model might
apply in small regions of parameter space.
According to one set of authors, the golden objects are those with low
$M$, high Eddington ratio, and very high
predicted temperature which could explain the Soft X-ray Excess
(though that doesn't seem very satisfying since
that feature is generic).  According to the other group of theorists,
the best hope lies in objects in the opposite
part of parameter space:  extremely massive black holes with low
Eddington ratios, and predicted very cool disks.  Neither group claims
the quasistatic model has relevance for the vast majority of objects.

Yet characteristic values and scaling relations based on the quasi-static
disk model are still
routinely used by most authors, generally without apology.  Remember
that the observations
don't just rule out quasi-static  models (the only ones with
predictive power!):  the sizes rule
out energy release {\it following the gravitational potential well}
expected for all black holes.  That means we aren't even close to
having the correct physics.

\subsection{Secrets of the X-ray Spectrum:  Mapping out the Kerr potential
           {\it Dubious}, 1990s--present}

In 1995, because of the advancement in X-ray spectroscopy from space
represented by the ASCA mission, it became possible to study the
emission line profiles in the brightest Seyfert 1s.

The strongest spectral feature in thermal AGN is the Fe K$\alpha$ line, which
has a rest-frame energy of 6.4 keV if it's neutral, and up to 6.9 keV
if it's highly ionized.
A discovery [with antecedents of course] was announced by Tanaka et al
in 1995 that almost everyone
(including me) thought was a breakthrough we were all waiting for.
This line was reported to be neutral and very broad in MCG 6-30-15,
of order 100,000 km/s, with a
particularly extended red wing to the {\it apparent} profile assuming no ``warm
absorption."  A (very
noisy) ``horn" appeared on the
red side of the profile, which would be suggestive of a disk origin.

Such profiles were interpreted as indicating the effects of Doppler shifts from
very fast motions and gravitational redshifts of the gas. This
suggests that they are produced very close to the putative supermassive black
holes.  In particular, it was argued that this line is produced by external
illumination of relatively cool gas, arranged in an accretion disk, by
a (mysterious but real) X-ray continuum source hovering above the putative disk.
The far side is of course similar, but unseen.

To some extent in this scenario, we'd be mapping out the relativistic region of
the putative supermassive black hole.  This ultimate key region had never been
explored so directly before.

The cleanest test of this disk-reprocessing interpretation in my opinion is to
see whether the emission line luminosity responds in real time to changes in
the driving continuum luminosity. It took almost no time for some of
the same authors as those of
the Fe K$\alpha$ discovery paper, using the same exact 4 day long data
set, to falsify the prediction
(Iwasawa et al 1996), although with much special pleading they claimed
they could save the model.
Iwasawa et al showed that the published ``disk-like" [e.g. the noisy red horn]
spectrum averaged over the 4 day integration never actually existed at
any one time!  It was an {\it artifact} of the summing over the particular observing
interval, which was set not for a physical motivation but by
scheduling and competition from other proposals.

For many years before and after the discovery, all authors similarly
reported that the lines do {\it not} respond intelligibly to continuum
changes on short (light travel) timescales, as virtually required
by the models.  Only a very few critical thinkers seemed to care about
this cognitive dissonance.

Minuitti et al (2003) did care about the problem, and developed what I
consider to be an
epicycle, called the ``light bending model."  (It's called a model
though it has very little
physics in it.)  The model asserts that the hovering X-ray continuum
(called the corona) comes
not from a fixed height above the putative disk, the previous fiducial
scenario adopted to minimize free
parameters.  Instead the corona\footnote{Recent papers also invoke ad hoc
changes in the size of the corona, ``to preserve the phenomena."} can now
be raised and lowered as needed
to fit the data.  In particular, the ubiquitous rapid continuum
variability in thermal AGN was seen as illusory,
a result of rapid vertical motions of the ``corona," which would throw
a highly variable flux our way,
but conspire to deliver a relatively
constant driving luminosity to the K$\alpha$ emitter.  Again, the sole
purpose was to explain (away) the lack of line response
by attributing the continuum changes not to actual changes in power.
But the lines do vary, they just don't follow the continuum. Nevertheless
sufficiently violent and ad hoc vertical motions of the corona
could scramble the correlation between the driving
continuum and the K$\alpha$
line, and to scramble it so much that it can never be recognized.
This paper made some predictions\dots
I'm not aware of any claims that they came true, but to my surprise, most of the
community seemed to be satisfied that these authors ``solved" the
problem.  All you have to do is say, ``Light bending!"

Incidentally, the X-ray illumination must also move around above the
disk in the radial and azimuthal
directions in an ad hoc manner to produce the changing bumps and
wiggles in the observed profile (Iwasawa et al 1996).

Cooler heads such as  T J Turner, K Weaver, T Yaqoob, L Miller and several
others have produced plausible (if less exciting) explanations for the apparent
far red wings of the X-ray Fe K$\alpha$ line profile as largely a
spurious interpretation
of the effects of absorption from ionized gas in front of the
continuum sources.  Complex and variable ionized and neutral absorbing
matter is almost ubiquitous in AGN, and well documented. Miller et al (2009)
showed that even the discovery object, MGC 6-30-15, can be
very well fit this way, and it resolves the problem of the lack
of intelligible line response to an apparently changing continuum.

The highly ionized absorbing gas (the ``Warm Absorbers") are well studied and
very widespread.  When one concentrates
on the few ``clean" objects which happen to have weak absorption (clearly
the best thing to do), the crucial
far red wings no longer appear!  Some great examples were shown by Patrick et
al in 2011, who demonstrated that only modestly broadened lines are {\it allowed} by the
data (and not necessarily even needed).  Interestingly, the original
poster-child, MGC 6-30-15, is very ``dirty" in the sense that it has
strong, variable, and
complex absorption, and as noted, can be modeled with no broad K$\alpha$ line at
all (Miller et al 2009).

Now let's go back to the contentious issue of the response of the fluorescence
line to continuum changes.  Many of the same authors that found no
line response to continuum changes previously have changed their
analysis method and
now find that the lines {\it do} show this effect!

\vfill\break

In the optical regime, the study of the response of the broad emission
lines to continuum changes is
extremely well developed.  The standard and minimum acceptable
demonstration of line response requires:
1) a clearly shown line/continuum separation.  This is much {\it more}
important, but largely lacking, in the X-ray papers.  They very rarely
show figures which allow
the reader to form an opinion on the decomposition, and it's so
important because of the claimed
extreme breadth of the line.\footnote{Disk advocates almost never show the spectra
below 1--2 keV, so the reader can't form an opinion on the all-important line/continuum
decomposition.  They tend not to show unprocessed observations at all, but only the
data divided by various models. This would never fly in the optical community.  I
did notice a couple recent papers by authors not so wedded to the disk interpretation
which do show the actual observations, and it is very plain that the continuum
placement is highly subjective---there is really no obvious continuum seen between
the broad spectral features. These plots dramatically illustrate that ``warm
absorption" can cause the continuum placement to be too low in the 2--5 keV region,
producing a spurious long red wing to the K$\alpha$ line profile. As a random example
see Fig.~5 in Pons and Watson 2014 (although I'm not persuaded by the other conclusions of that
paper).}
2) light curves for the continuum and line flux;
3) a cross-correlation curve showing the line flux response over time.
It can be made for separate parts of the profile if the signal to
noise ratio is high.

X-ray astronomers claiming line responses to continuum changes never
(as far as I know) follow any of these precepts. These failings interact in a
pernicious way. For example, there are well documented complex interband
continuum phase lags in AGN X-ray spectra, so that even if a phase lag is
correctly inferred in the region of the putative far red K$\alpha$ wing
relative to the Fe-ionizing continuum, it might arise from the underlying continuum.
As examples, Legg et al (2012) and Miller et al (2010) should be consulted. 

The reason some astronomers find line responses today is that they have
adopted a new method with a lot more
freedom and subjectivity.   These authors now take only particular
Fourier components of the line and continuum
light curves, and find some lags by selecting the Fourier frequencies
{\it a posteriori}.  Time will tell if the newly reported short
lags are real.

Also the papers that that I know which report time lags look for phase
differences between the line and continuum at the same temporal frequency. But
the response should actually be smoothed out by light travel time across the
reprocessor, so as I understand it, the analyses aren't self-consistent. I
think it would be far more reliable to calculate the cross-correlation and just
display the response (``Transfer") function.  This could be done as a function
of energy within the profile, as optical astronomers do.

These papers may all be correct but disk fluorescence proponents seem
to invoke more and more epicycles
as the monitoring data improve.  If the lags inferred for the different
parts of the line profile are set by the geometry of a spinning
quasi-Keplerian disk,
shouldn't they stay about the same over time?  Please read Alston et
al.\/ 2013, Sec.~4, as an example of what can
really happen. Another example is discussed by Kara et al (2014). It's also
worth noting that extravagant abundances are sometimes required, e.g. 7--20x
solar M1H0707-493; see discussion in Done and Jin (2015).

Switching gears slightly, the very small radius of the innermost stable circular
orbit for rapidly rotating black holes has led to a substantial literature
running the argument
backwards:  a highly redshifted wing, for which the observed
line energy is only $\ltwid$ half the rest energy, requires a rapidly spinning hole.
So there are many fits to line profiles which people use to infer
black hole spin. In fact this is now a huge and widely accepted industry.
To get an idea of the robustness of spins inferred from X-ray spectra, please
read Sec.~4.3.4 from Patrick et al. (2012) on the spin of the Seyfert NGC 3783.
In the previous paper (Patrick et al 2011) of this outstanding series of three,
the authors analyzed a magnificent 210 ksec Suzaku observation, placing an
upper bound of 0.31 on the spin parameter. The exact same data were analyzed in
Brenneman et al (2011); those authors could say with 90\% confidence that the value
is greater than 0.98.

The situation seems little better for X-ray binaries. For example, recent papers
on Cyg X-1 have reported values of $0.05\pm0.01$ and $0.97\pm0.014/-0.02$,
where even the asymmetry of the tiny errors has been quantified.

Another puzzle in the disk interpretation: they are undetected at a constraining
level in {\it most} AGN, yet are attributed to the essential relativistic accretion
process thought to provide {\it all} of the energy to thermal AGN generically.

Now let's try to combine the ``great discoveries" of the optical/UV emitting
accretion disk and the spin-measuring Fe K$\alpha$ profiles.

The (falsified) standard thermally emitting optical/UV-emitting accretion
disk gets the bulk of its energy from basically the
same relativistic
region where the Fe line arises.  Yet there is almost no known empirical connection
between the emitting disks and the flourescing disks, which are considered as
purely passive reprocessors in all X-ray modeling that I've seen. {\it Yet
close connections are expected.} For example, most fluorescence models, and
{\it all} those reporting spins, require that the disks extend as optically thick
geometrically thin structures to only a few gravitational radii, in particular
to the innermost stable circular orbit. The inner edges of such disks should be
very hot, and this should be manifest as very blue UV/FUV continua, yet this is
virtually never seen. In fact, observation-oriented theorists trying to fit
optical/UV spectra to accretion disk models must truncate the disks well
outside this region (e.g. Jin et al 2012; Done et al 2012; Laor and Davis 2014)!

Again, relatively few astronomers seem to be bothered by the cognitive dissonance.

\section{A few observations and calculations which could address these
problems fundamentally}

\subsection{Advanced X-ray reverberation mapping} Really robust and detailed line/continuum response studies with future
advanced X-ray telescopes will tell us a lot about what we really want to know: the
spacetime geometry and mass flows in the innermost regions of the highly warped
spacetime extremely close to the black holes.

\subsection{Getting the true central-engine optical/UV spatial energy distribution}
We observers owe the theorists real spectral energy distributions of the
energetically dominant ``central engine" radiation, that is, without
the daunting contributions from
various kinds of reprocessed emission.  In some regions of the optical/UV spectra,
small nearly-pure continuum windows are available, but in large wavelength intervals
this is not the case.\footnote{Contamination of the intrinsic continuum spectrum by atomic and dust emission can be fierce.
The spectra in Stevans et al 2014 illustrate the situation in the far-UV; see e.g. Vanden Berg et al 2001 for the
near-UV to optical region.  Past one micron, with rare exceptions, nothing can be 
learned about the ``central engine" spectrum at all without polarimetry because the entire region
is heavily dominated by dust emission.} Yet it is possible to measure the true continuum emission
in some cases using polarimetry. Small but very successful beginnings at discovering the true shape
and spectral features of the Big Blue Bump have been published by Kishimoto et al 2004, 2008 and references therein.  We have
found that in certain quasars, the optical
continuum comes to us with a small but detectable electron-scattered fraction,
and it is thus slightly polarized; this is physically similar to the
periscope affect
discussed above in the context of Unified Models, but on scales of
order a million time smaller.  It's a
fantastic piece of luck because the contaminating components (starlight from
the host galaxy, atomic emission lines and bound-free continua, and infrared
radiation from glowing solid particles) are sometimes all unpolarized. So we need only
plot the polarized flux spectrum to see the {\it isolated} central engine spectrum!
We've discovered the first spectral feature intrinsic to the actual AGN engines
in this way, absorption in the Balmer continuum (Kishimoto et al 2004). We also know that
there is a sharp slope change longward of $1\mu$ (Kishimoto et al 2008): the
central engine spectra are much bluer in the near-IR than in the optical, with
spectral index $\sim+0.3$, when the dust emission is removed with polarimetry.
Later several objects were found with intrinsically weak near-IR dust emission,
and this blue slope was confirmed in total flux.

\subsection{Advanced numerical simulations}
Very detailed relativistic magnetohydrodynamic simulations of matter flowing
into a black hole, with the inclusion of such essentials as
dissipation and radiation will be required.
Software breakthroughs, aided by Moore's law, are starting to achieve
some traction in certain special cases. So far we have only explored this method
down to around 3000\AA\ in the rest frames, at high SNR, but we have taken data
pushing down through and past the Small Blue Bump covering 2000-4000\AA\ to
find the shape in this heavily contaminated region, and we are also exploring
the Ly continuum region where very surprising but tentative results have been
reported.  It goes without saying that this is incredibly difficult.

\subsection{The Green's Functions for AGN: Tidal Disruption Events}
One thing we know about AGN is that they are a complicated mess.  Theorists
need a simple well-defined problem with known initial conditions.
After M Rees whipped
us into a frenzy of excitement about it in 1988, many of us have
waited for the day
when we could observe isolated starved black holes, and then throw a
single simple piece of
matter (a star!) into them, sit back and see what happens!  I call the
various resultant displays the ``Green's
Functions" for supermassive black hole accretion.  Such events have been
persuasively identified by S Komossa and others --- the data gathered
during the events so far is limited, but this is improving rapidly.  Will we see
short-lived quasars
when this happens?  Feast your eyes on recent papers such as Yang et al
2013 and Arcavi et al
2014.  This is the most exciting thing going on in the AGN field today.

\section*{Acknowledgements} I benefitted by advice from various people,
including Jane Turner, Martin Gaskell, Eric Agol, Jason Dexter, Chris Done,
Ari Laor, Sebastian Hoenig, Todd Hurt, Patrick Ogle, and Rob Geller.
\vspace{-5mm}
\section*{References}

\nocite{*}
\bibliographystyle{apsrev4-2}
\bibliography{agnpuzzle}

\section*{Appendix: Some poor practices in AGN research}

For those with a taste for spleen, I present a list of some practices which are almost generic
in AGN research, but which seem erroneous to me. This is an idiosyncratic list, written using
the stream-of-consciousness technique. Please let me know where I've gone wrong.

1. Vast amounts of observing time have been spent on spectra of AGN and infrared
galaxies, at telescopes with polarimetry modules that the observers didn't both
to place in the beam.  Yet look at the value this often adds.  For example,
observers wasted a lot of Keck time taking spectra of Cygnus A in search of weak broad emission lines.
None were detected, and tight (but spurious) upper limits were claimed.  Please look at Ogle et al 1997, Figs.~1 and 2,
which resolved the situation entirely.  Similarly thousands of quasar spectra have been taken in total
flux, though polarization could have been taken almost for free.  As noted earlier, Kishimoto et al (2008) used
the polarized flux of some clean objects to get to the intrinsic spectrum of the central engine.  
A huge amount of work has been done on the host galaxies of Type 1 quasars in total flux mode, with crippling
contamination by nuclear light.  This is despite the fact that we know (e.g. from spectropolarimetry)
that Type 2 quasars are the same objects, but with Nature's coronagraph completely blocking the nuclear
light.\footnote{As far as I know, only one group has taken advantage of Nature's coronagraph to study
quasar host galaxies. Zakamska et al (2006) rightly state that ``spectacular, high-quality" host images
were obtained.  Also see their spectropolarimetry revealing the hidden quasars in Zakamska et al 2005.}

Infrared galaxies often have spectacular polarization properties.  Take a look at G.~Schmidt (1975)'s image of M82,
achieved with a 2.4m telescope, photographic plates (!!), and 12 minutes of exposure time per waveplate position.
The extremely high polarization of the lines as well as the continuum ($\sim25$\%) means that much of the emission is not
produced in situ, but it has been analyzed as if it were.  One could also observe high latitude clouds to find 
the spectrum of M82 ``as seen from above."  The dust can also provide an approximate but robust estimate for
the cool-phase mass, a key ingredient in galactic wind models.  We (Hoenig et al) have gathered some great polarization images of other infrared galaxies. 
At the high luminosity end, all of the original ``Hyperluminous" infrared galaxies show quasars in polarized flux.
It's too bad people spent all that time on total flux spectroscopy. Of course a high-ionization infrared galaxy is just
another name for a luminous Type 2 AGN.  

In order to get the polarization at Keck, you have to type ``POLAR $<$cr$>$." 
There are no plans for polarization analyzers on any of the big new telescopes being built, because no one uses
the instruments on the current telescopes.

2.  Plots of one luminosity measure vs.\/ another within a class of objects almost always find good
correlations, because big things are bigger. Consider a plot of the number of bookstores vs.\/ the number
of bars in cities and towns across America. I'd expect a great correlation over many orders of
magnitude, with perhaps modest dispersion and essentially infinite statistical significance. Everyone knows
that doesn't mean that readers like to drink, or that drinkers like to read.
I'm tempted to give an example of a very influential paper from 1991 which used this method to arrive
at an erroneous conclusion that took a decade or two to fuzzily reverse.
The key reference illustration here is Kennicutt (1996), where he shows a plot of CO vs.\/ far-IR luminosity 
for galaxies, but then to extend the baseline he adds a burning cigar, a Jeep Cherokee Wagon,
the 1988 Yellowstone park forest fire, Venus, and the observable universe.

It's helpful to plot the fluxes as well as the luminosities. Flux plots are also dangerous, but the horrible biases
introduced are {\it different} from those with luminosity, so when the correlation shows up in both,
I get a warm feeling. Note that I am by no means an expert in statistics.

3.  Another extremely common type of plot involves A vs.\/ B/A or some other mathematically dependent or partially
dependent quantities. 
The dependence is often hidden because one or both observables are relabeled as physical quantities
based on some dodgy model. Of course if A and B are unrelated, you are guaranteed a spurious negative
correlation, if there is a finite amount of intrinsic dispersion or observational error. The significance
will be arbitrarily high if you have a lot of objects. Egg-shaped distributions are especially suspect.
Even if a ``real" correlation is present, the bias contributes to the apparent correlation, so that
the slope isn't meaningful in general, yet such slopes are routinely quoted and used.

There are some plots, I think those with narrow low-dispersion features and only very
mild dependences of the axes, for which this bias isn't important. An example is
the HR Diagram. But diagrams of $B-V$ vs.\/ $V$ should really use some kind of
average of $B$ and $V$ for the dependent variable.

4.  The Gaussian function is extremely pathological because the far wings are vanishingly small.
(Think of the Curve of Growth.) The distribution involves the exponential of the
square of a deviation from an apparent correlation line. Since outliers are
impossibly unlikely, in such a distribution, you can rule out a null result
with one object. Few quantities used by astronomers are Gaussian-distributed.
It works for thermal noise, counting photons, etc,
but for few if any actual population distributions of macroscopic objects. 

5.  How do astronomers treat limits in null and correlation tests? 
To use limits ``correctly", a survival function must be invoked and rationalized.
Astronomers arbitrarily pick something like Proportional Hazards 
(e.g. the Kaplan-Meier method) without any basis, or even comment. These approaches,
derived in the context of actuarial tables\footnote{*Mathematicians are a rigorous
lot. Wiki says, ``The survival function is usually assumed to approach zero as
age increases without bound, i.e., $S(t) \to 0$ as $t \to\infty$, although the
limit could be greater than zero if eternal life is possible."}, have no obvious
applicability to astronomical data sets, which depend on luminosity, flux,
or other distributions which are {\it a priori} unknown.

6.  As noted, many X-ray astronomers in particular decline to plot fully
calibrated data covering the entire observed energy range. Instead they show count
spectra, spectra without enough coverage of the ``continuum" to judge the
believability of the Fe K-$\alpha$ far wings, or most appallingly, {\it spectra
divided by models}! I'd like to see an optical paper showing only spectra divided by Cloudy models.
These authors are to some extent selling models, and the process reminds me of
an actual Freudian slip: ``I'll see it when I believe it."

7.  It's still extremely common for people to mis-state the meaning of null test
statistics. For example, ``The thick lines indicate where a Welch's $t$ test
shows the mean SEDs have less than a 1\% chance of being the same ($p < 0.01$)."
(I read this aloud to a particle physicist analyzing LHC data because he happened
to be in my office and he burst out laughing.)

No, the test shows that uncorrelated data would, in a certain sense, look as
correlated as yours (in your chosen test statistic) only 1\% of the time.  This is
certainly not a semantic issue. In my critical thinking lesson for the Astro 1 Honors section, I show a stupid
(old) article from a financial magazine which noted that over the previous 10 years, two methods of predicting
the next year's stock market performance worked equally well, getting the correct result 9 times. One involved
insider trading: lots of buys portend a rising market. The other method was based on whether an American
Football League or a National Football League team won the championship. The statistical significance is the same.
Then I ask which method they would use for investing their retirement account funds and they get it right away.

Steve Reynolds once told me the First Theorem of Metastatistics:
``Half of all 3-sigma results are true."  (He's since raised it to 5-sigma.)  And here is an apocryphal quotation
attributed to Martin Rees: ``It would be really funny if nothing funny ever happened."

8.   In comparing samples to see whether or not they may differ only in
orientation, a mighty torrent of negative results were reported for many years,
both in the context of the beam model, and then all over again for the torus
model. It was pointed out that e.g. the Seyfert 1 prototype, NGC 4151, has much
weaker emission in the IR, in CO, etc, compared with the Seyfert 2 prototype
NGC1068. Therefore they can't differ only in orientation. Only trouble is, if we
select by something like UV excess, which is comparable in the two objects, we
are comparing the actual luminosity of NGC 4151, to the 1\% of the luminosity
which is scattered into the line of sight for NGC 1068. So NGC 1068 is 5
magnitudes higher on the luminosity function. No wonder it has more of everything.
In the history of testing the beam model, properties of superluminal sources
were compared with those of big luminous FR II doubles; since the former were
selected on the beamed emission, they turned out to have much lower diffuse
lobe emission, way down in the FR I range in most cases. You must select your
sample on some hopefully nearly isotropic property of the AGN. It does {\it not} work
to select on the host galaxies.

9.  What is the AGN torus? Joe Miller and I didn't introduce that term, and
didn't draw an outer boundary for the equatorial obscurer, because we had no
information on that. The torus is historically defined to be the thing that
causes {\it optical} nuclear photons to escape preferentially in the polar
direction, the polarization position angle having been found to be perpendicular
to the radio axes in Type 2 objects. The anisotropy is very substantial because
the percent polarization of the scattered light is usually relatively high (at
least by selection). Unspecified is the radius at which the shadowing occurs,
except that it must be inside the narrow line region. But it's likely that they
are quite messy: see what I call the (very) ``Sloppy Torus" CO source in NGC 1068,
Garcia-Burillo et al 2014; it looks most torus-like in Fig.~4. The dynamics are already
known to be complex. In some cases the torus may just be a dusty wind. The fast growing field of
infrared interferometry will tell us a lot, as will the very sensitive and sharp ALMA millimeter array.

One should never refer to the size of the torus without stating exactly what you mean for this
edge-darkened source. It does seem clear though that essentially every thermal AGN has
a significant covering factor of dust at the sublimation radius, so the inner boundary is
relatively well constrained.

10.  There is an industry finding, studying and theorizing about objects which are like
Seyfert 1 nuclei, but ``lack broad lines."  I urge anyone interested in this literature
are urged to read my critical comments in Sec.~2 of Antonucci 2012. Few of the preferred
candidates are strong. I'll take this last chance to tout that long detailed review
of unification and central engine types, which carefully covers the data at all wavelengths.

11.  The mainframe (!) computer went down one day when I was a young postdoc at STScI. All the astronomers
were huddled in the library, reading papers and discussing science instead of pecking away at their terminals.
Visiting Santa Cruz, California, I saw a bumper sticker with the wisdom of that fair city: ``Don't just do something, 
sit there."

12. A great many papers refer to the peak of a spectrum and the corresponding
temperature for a blackbody emitter. But there is no consistency on how the
peak is defined.  In fact it's not usually even specified.

The solar spectrum is said to peak at a wavelength of $5 \times10^{-7}$m, which
comes from Wien's blackbody law, $\lambda_{\rm max} = 2.9\times10^{-3}$ m $K/T$.
This applies only to a plot of flux (or intensity) per unit wavelength interval,
$F_\lambda$, and makes objects seem ``bluer" than they really are.  This is easy
to see:  many astrophysical objects emit strongly in the x-ray and gamma ray.
In a plot of flux per nanometer, for example, the entire hard X-ray and gamma ray
luminosity would fall into a single bin, the one going from 0 nm to 1 nm! Plots
of the flux per unit frequency are similarly flawed, but in the opposite sense.

When plotting a wide band spectral energy distribution, almost all astronomers
now use $\nu F_\nu$ or, equivalently, $\lambda F_\lambda$.  This is proportional
to flux per unit logarithmic frequency or wavelength interval, or more simply,
flux per decade. This shows the actual distribution of flux across the spectral
energy distribution. The equivalent of Wien's law when plotting $F\nu$ is
approximately $\lambda_{\rm max} = 5.1\times10^{-3}\ {\rm m K/T}$, which
corresponds to an apparent spectral peak of
$\nu_{\rm max} = (5.9\times10^{10}\ {\rm Hz/K) T}$. When plotting the recommended
$\nu F_\nu$, the peak $\lambda_{\rm max}$ occurs at $3.7\times10^{-3}\ {\rm m K/T}$
and $\nu_{\rm max}=8.2\times10^{10}\ {\rm T(Hz/K)}$.



\end{document}